\documentstyle[12pt,thmsa,a4,sw20lart]{article}
\input tcilatex
\QQQ{Language}{
American English
}

\begin{document}

\title{Spectrum of surface-mode contributions to the excitation probability for
electron beam interacting with sharp-edged dielectric wedges}
\author{H.B.Nersisyan and A.V.Hovhannisyan \\
%EndAName
Institute of Radiophysics and Electronics,\\
Ashtarak-2, 378410, Armenia}
\maketitle

\begin{abstract}
The interaction of a nonrelativistic charged particle beam, travelling
parallel to the surface of a sharp-edged dielectric wedge is analyzed. The
general expressions for excitation probability are obtained for a beam
moving along the direction of a symmetry axis, either outside or inside the
dielectric wedge. The dielectric function of the medium is assumed to be
isotropic, and numerical results are given for the materials of experimental
interest.
\end{abstract}

\section{Introduction}

The tradition of analyzing material targets from the energy-loss
spectroscopy of charged particles scattered through or near a scatter has
continued to enrich physics over the past several decades. Applications have
been found in nuclear and particle physics, atomic and molecular physics,
and in condensed-matter physics. Recently, biological physics has benefited
as well, particularly from the use of electron-energy-loss spectroscopy
(EELS) performed using scanning-transmission-electron microscopes (STEM).
Using a STEM one can obtain information on the size, shape, composition, and
location of isolated particulates embedded in a host material (composite)
and thus obtain three-dimensional chemical maps with high resolution (Chen 
{\it et al} 1986).

In a typical STEM configuration, a well-focused 0.5-nm probe of 50-100 keV
electrons provides a high-resolution transmission scanning image for samples
with complex structures. It also yields from selected regions of the
structure, x-ray emission spectra and electron energy-loss spectra.
Quantitative theories have been developed to analyze the experimental
energy-loss spectra in some simple cases.

Solutions, within the classical dielectric theory, have been worked out for
a number of cases involving planar interfaces (Echenique and Pendry 1975),
spheres (Ferrel and Echenique 1985, Echenique {\it et al} 1987), cylinders
(Walsh 1989, Zabala {\it et al} 1989), spheroids (Illman {\it et al} 1988),
and parabolically shaped wedges (Garcia-Molina {\it et al} 1985). For these
simple geometries, experimental results show that dielectric excitation
theory is capable of predicting energy-loss spectra, allowing a fully
consistent dielectric characterization of an interface or a small particle
(Walls and Howie 1989, Rivacoba {\it et al} 1992).

In this paper we focus our attention on the calculation of excitation
probability for point charge particle moving parallel to the sharp-edged
dielectric wedge whose boundary is formed by the intersection of two
semi-infinite planes making an interior angle of $\alpha $. Our interest is
explained by some experiments that were reported in (Marks 1982, Cowley
1982, Wheatley {\it et al} 1983). The targets ($MgO$, $NiO$, $Al_2O_3$,
etc.) bombarded in the experiments are of cubic symmetry (i.e. have
sharp-edged form), about 20-200 nm in size, and the electron beam is
oriented along the principal crystallographic directions. Marks (1982) have
measured the spectra of energy-losses of electron beams interacting with
small crystallites of $MgO$. The calculations for excitation probability
were done considering the crystallites as semi-infinite surface related to
electron beams. Using the classical theory of impact-parameter-dependent
energy losses for planar interfaces Wheatley {\it et al} (1983) applied the
planar results for spheres while resolving the force in the track direction.
Garcia-Molina {\it et al} (1985) calculated the energy-loss function and the
excitation probability of the wedge modes due to a pointlike electron beam
moving parallel to the dielectric wedge surface with a parabolic-cylinder
boundary.

Here we shall calculate the excitation probability of sharp-edged modes, due
to an electron beam passing parallel to the dielectric wedge with a local,
but otherwise arbitrary, dielectric function $\varepsilon (\omega )$. We
shall limit our calculations to the nonretarded limit. The limiting cases
for the potential and excitation probability are shown. First, from exact
expressions for excitation probability we derive the expressions for planar
geometry in the limiting case of $\alpha =\pi $. Second, in Appendix we also
show that in the static limit ($u\rightarrow 0$, where $u$ is the velocity
of particle) from our results for the potential follows the expression for
conducting sharp-edged wedge ($\varepsilon \rightarrow \infty $) (Landau and
Lifshitz 1982). In Sec. III we have utilized available bulk optical data
(Roessler and Walker 1967) to calculate the differential energy-loss
probability for $MgO$. The theoretical predictions are compared with
experimental data, and it is shown that the main features of the experiments
reported by Marks (1982) can be explained by the theory. Also, our results
are compared with a parabolically shaped wedge (Garcia-Molina {\it et al}
1985). In Sec. IV we present our conclusions and comments on the results and
discuss opportunities for possible future work.

\section{Energy loss and excitation probability}

Take dielectric wedges whose boundaries are formed by the intersection of
two semi-infinite planes making an interior angle of $\alpha $ infinite in
the $z$ direction. Let in the cylindrical system of coordinates $\rho ,$ $%
\theta ,$ $z$ the azymutal angle $\theta $ is measured from one of wedges
sides. We consider two media filling the spaces $-\infty <z<+\infty $, $%
0\leq \theta \leq \alpha $, $\alpha \leq \theta \leq 2\pi $, and
characterized by an isotropic dielectric functions $\varepsilon _1(\omega )$
and $\varepsilon _2(\omega )$ respectively (see figure 1). The incident
charged particle with the charge $q$ moves with a velocity $u$ directed
along the $z$ axis and has the following coordinates $\rho =a,$ $\theta
=\gamma $ ($\gamma <\alpha $), $z=ut$.

The study of the electrostatic edge modes along a sharp-edged wedge is due
to Dobrzynsky and Maradudin (1972), who solved Laplace's equation in the
appropriate coordinate system. Davis (1976) has considered the electrostatic
modes of a hyperbolic cylinder and has concluded that the results in the
work of Dobrzynsky and Maradudin (1972) are associated with the sharpness of
the edge of the wedge. Here we shall only give the main steps in the
derivation of the electrostatic potential originated as a result of an
electron beam traveling along the wedges surfaces.

We solve Poisson's equation for the potential

\begin{equation}
\nabla ^2(\widehat{\varepsilon }\varphi )=-\frac{4\pi q}a\delta (\rho
-a)\delta (\theta -\gamma )\delta (z-ut),
\end{equation}
where the charge density associated with an electron beam, which is
described classically by a $\delta $ functions, $\widehat{\varepsilon }$ is
the operator of dielectric permeability of the medium (Landau and Lifshitz
1982).

It is convenient to work in Fourier space:

\begin{equation}
\varphi (\rho ,\theta ,\xi )=\int_{-\infty }^{+\infty }d\omega \exp (i\omega
\xi /u)\varphi _\omega (\rho ,\theta ),
\end{equation}
where $\xi =z-ut$. Then, Poisson's equation becomes, in cylindrical
coordinates, with $\varphi _\omega (\rho ,\theta )$,

\begin{equation}
\left( \frac{\partial ^2}{\partial \rho ^2}+\frac 1\rho \frac \partial
{\partial \rho }+\frac 1{\rho ^2}\frac{\partial ^2}{\partial \theta ^2}-%
\frac{\omega ^2}{u^2}\right) \varphi _\omega (\rho ,\theta )=-\frac{2q}{%
au\varepsilon (\omega )}\delta (\rho -a)\delta (\theta -\gamma ).
\end{equation}
The solutions for the electrostatic potential in the regions $0\leq \theta
\leq \alpha $ and $\alpha \leq \theta \leq 2\pi $ are respectively

\begin{eqnarray}
\varphi _\omega (\rho ,\theta ) &=&\frac q{\pi ^2u}\int_{-\infty }^{+\infty
}d\mu K_{i\mu }(ka)K_{i\mu }(k\rho )\times \\
&&\times \left\{ 
\begin{array}{l}
\frac 1{\varepsilon _1(\omega )}\left\{ {\rm ch}\left[ \mu \left( \pi
-\left| \theta -\gamma \right| \right) \right] +\psi _{1\omega }(\mu ,\theta
)\right\} ,\quad 0\leq \theta \leq \alpha \\ 
\frac 1{\varepsilon _2(\omega )}\psi _{2\omega }(\mu ,\theta ),\quad \alpha
\leq \theta \leq 2\pi .
\end{array}
\right.  \nonumber
\end{eqnarray}
where $K_\nu (x)$ is a modified Bessel function of order $\nu =i\mu $, $%
k=|\omega |/u$,

\begin{eqnarray}
\psi _{1\omega }(\mu ,\theta ) &=&A_\omega (\mu ){\rm ch}(\mu \theta
)+B_\omega (\mu ){\rm sh}(\mu \theta ), \\
\psi _{2\omega }(\mu ,\theta ) &=&C_\omega (\mu ){\rm ch}(\mu \theta
)+D_\omega (\mu ){\rm sh}(\mu \theta ).  \nonumber
\end{eqnarray}

The first term in the large brackets of equation (4) corresponds to the
potential of the particle in the unbounded medium with dielectric function $%
\varepsilon _1(\omega )$ (particular solution of the inhomogeneous equation
(3)). The other terms in equation (4) correspond to the potential generated
due to the existence of the interfaces (solution of the homogeneous equation
(3) or Laplace's equation (Dobrzynsky and Maradudin 1972)).

The coefficients $A_\omega (\mu )$, $B_\omega (\mu )$, $C_\omega (\mu )$ and 
$D_\omega (\mu )$ are determined by the boundary conditions. The potential
must be continuous at $\theta =0$ and $\theta =\alpha $, and the normal
component of the Fourier amplitude of the electric displacement vector must
be continuous at $\theta =0$ and $\theta =\alpha $. From equation (4) we can
obtain the set of following equations for coefficients $A_\omega (\mu )$, $%
B_\omega (\mu )$, $C_\omega (\mu )$ and $D_\omega (\mu )$

\begin{eqnarray}
&&\frac 1{\varepsilon _1(\omega )}\left[ {\rm ch}[\mu (\pi +\gamma -\alpha
)]+A_\omega (\mu ){\rm ch}(\mu \alpha )+B_\omega (\mu ){\rm sh}(\mu \alpha
)\right]  \nonumber \\
&=&\frac 1{\varepsilon _2(\omega )}\left[ C_\omega (\mu ){\rm ch}(\mu \alpha
)+D_\omega (\mu ){\rm sh}(\mu \alpha )\right] ,
\end{eqnarray}

\begin{equation}
\frac 1{\varepsilon _1(\omega )}\left[ {\rm ch}[\mu (\pi -\gamma )]+A_\omega
(\mu )\right] =\frac 1{\varepsilon _2(\omega )}\left[ C_\omega (\mu ){\rm ch}%
(2\pi \mu )+D_\omega (\mu ){\rm sh}(2\pi \mu )\right] ,
\end{equation}

\begin{eqnarray}
&&-{\rm sh}[\mu (\pi +\gamma -\alpha )]+A_\omega (\mu ){\rm sh}(\mu \alpha
)+B_\omega (\mu ){\rm ch}(\mu \alpha )  \nonumber \\
&=&C_\omega (\mu ){\rm sh}(\mu \alpha )+D_\omega (\mu ){\rm ch}(\mu \alpha ),
\end{eqnarray}

\begin{equation}
{\rm sh}[\mu (\pi -\gamma )]+B_\omega (\mu )=C_\omega (\mu ){\rm sh}(2\pi
\mu )+D_\omega (\mu ){\rm ch}(2\pi \mu ).
\end{equation}
Only $A_\omega (\mu )$ and $B_\omega (\mu )$ are of interest as they
represent the coefficients of the homogeneous portion of the potential and
thus are needed to obtain the self-energy and stopping power.

We seek the dissipative component of the force acting on the beam moving
near the wedge surface. We neglect quantum recoil effects and assume that $u$
is constant (i.e., the external charge acts as an infinite source of energy
and momentum). The negative of the dissipative component of the induced
force is the specific energy loss (Ritchie 1957)

\begin{equation}
-\frac{dW}{dz}=\int_0^\infty d\omega \hbar \omega P(\omega ),
\end{equation}
where

\begin{equation}
P(\omega )=\frac{2q^2}{\pi \hbar u^2}\left\{ {\rm Im}\frac{-1}{\varepsilon
_1(\omega )}\ln \frac{k_cu}\omega +\frac 2\pi \int_0^\infty d\mu K_{i\mu
}^2\left( \frac \omega ua\right) {\rm sh}\left[ \mu \left( 2\pi -\alpha
\right) \right] Q_\omega (\mu )\right\}
\end{equation}
is the excitation probability,

\begin{equation}
Q_\omega (\mu )=-{\rm Im}\left\{ \frac{\eta _s(\omega )}{\varepsilon
_1(\omega )}\frac{\eta _s(\omega ){\rm sh}\left[ \mu \left( \pi -\alpha
\right) \right] +{\rm sh}(\pi \mu ){\rm ch}\left[ \mu (2\gamma -\alpha
)\right] }{{\rm sh}^2(\pi \mu )-\eta _s^2(\omega ){\rm sh}^2\left[ \mu
\left( \pi -\alpha \right) \right] }\right\} ,
\end{equation}
\begin{equation}
\eta _s(\omega )=\frac{\varepsilon _1(\omega )-\varepsilon _2(\omega )}{%
\varepsilon _1(\omega )+\varepsilon _2(\omega )},
\end{equation}
is the surface response function for plane geometry, $k_c=2mu/\hbar $ is a
cutoff wave number (Brandt {\it et al} 1974). Note that this definition of $%
k_c$ is valid when charged particle beam is sufficiently fast, i.e. $%
u>e^2/\hbar \simeq 2.2\times 10^8$cm/sec. As expected, the excitation
probability in equation (11) contains terms corresponding to both the
excitation of the bulk (the term ${\rm Im}(-1/\varepsilon _1)$) and of the
surface modes of the wedge. Therefore, if one wishes to probe the surface
excitation field, without interference from the bulk modes, the probe has to
be kept external to the wedge, as was done in the experiments (Marks 1982,
Cowley 1982, Wheatley {\it et al} 1983). One can calculate the specific
energy loss from equations (11) and (12).

The dispersion relation for the surface modes is the result of equating the
denominator in equation (12) to zero. As it follows from the expression
(12), there are two type of surface modes. The frequency of the first one
(so called even mode), in the case of $\varepsilon _2(\omega )=1$, $%
\varepsilon _1(\omega )=\varepsilon (\omega )$, is determined from the
dispersion equation:

\begin{equation}
\varepsilon (\omega )=-\frac{{\rm th}\left[ \mu \left( \pi -\frac \alpha
2\right) \right] }{{\rm th}\left( \frac{\mu \alpha }2\right) }.
\end{equation}
The electric potential in this surface mode is symmetric as related to the
symmetry plane of the wedges (the plane $\theta =\alpha /2$) (Dobrzynsky and
Maradudin 1972). The frequency of the second one (so called odd mode) is
determined from the dispersion equation:

\begin{equation}
\varepsilon (\omega )=-\frac{{\rm th}\left( \frac{\mu \alpha }2\right) }{%
{\rm th}\left[ \mu \left( \pi -\frac \alpha 2\right) \right] }.
\end{equation}
The electric potential in this type of surface mode is antisymmetric as
related to the symmetry plane of the wedges (Dobrzynsky and Maradudin 1972).
When $\varepsilon _2(\omega )=\varepsilon (\omega )$, $\varepsilon _1(\omega
)=1$, the above mentioned dispersion equations for the odd and even surface
modes change their places.

We consider first the case of a beam travelling external to the wedge, in
vacuum ($\varepsilon _1(\omega )=1$, $\varepsilon _2(\omega )=\varepsilon
(\omega )$). Then from equations (11)-(13) we have

\begin{equation}
P(\omega )=\frac{4q^2}{\pi ^2\hbar u^2}\int_0^\infty d\mu K_{i\mu }^2\left(
\frac \omega ua\right) {\rm sh}[\mu (2\pi -\alpha )Q_\omega (\mu ),
\end{equation}

\begin{equation}
Q_\omega (\mu )={\rm Im}\left\{ \eta (\omega )\frac{\eta (\omega ){\rm sh}%
\left[ \mu \left( \alpha -\pi \right) \right] -{\rm sh}(\pi \mu ){\rm ch}%
\left[ \mu (2\gamma -\alpha )\right] }{\eta ^2(\omega ){\rm sh}^2\left[ \mu
\left( \alpha -\pi \right) \right] -{\rm sh}^2(\pi \mu )}\right\} ,
\end{equation}
where

\begin{equation}
\eta (\omega )=\frac{\varepsilon (\omega )-1}{\varepsilon (\omega )+1}.
\end{equation}

We consider now the case of a beam travelling through the wedge, but
parallel to the edge. (i.e. $\varepsilon _2(\omega )=1$, $\varepsilon
_1(\omega )=\varepsilon (\omega )$)

\begin{equation}
P(\omega )=\frac{2q^2}{\pi \hbar u^2}\left\{ {\rm Im}\frac{-1}{\varepsilon
(\omega )}\ln \frac{k_cu}\omega +\frac 2\pi \int_0^\infty d\mu K_{i\mu
}^2\left( \frac \omega ua\right) {\rm sh}\left[ \mu \left( 2\pi -\alpha
\right) \right] Q_\omega (\mu )\right\} ,
\end{equation}

\begin{equation}
Q_\omega (\mu )={\rm Im}\left\{ \frac{\eta (\omega )}{\varepsilon (\omega )}%
\frac{\eta (\omega ){\rm sh}\left[ \mu \left( \pi -\alpha \right) \right] +%
{\rm sh}(\pi \mu ){\rm ch}\left[ \mu (2\gamma -\alpha )\right] }{\eta
^2(\omega ){\rm sh}^2\left[ \mu \left( \pi -\alpha \right) \right] -{\rm sh}%
^2(\pi \mu )}\right\} .
\end{equation}
Now both terms in equation (19) contribute to the specific energy loss of
the beam travelling through the wedge.

It is instructive to derive from equations (16) and (19) the excitation
probability quoted by Marks (1982) for an electron beam traveling parallel
to a semi-infinite dielectric occuping the region $0\leq \theta \leq \pi $
(or $\pi \leq \theta \leq 2\pi $), and at a distance $d$ from its surface.
Substituting $\alpha =\pi $ in the equations (16) and (19) we find

\begin{equation}
P(\omega )=\frac{4q^2}{\pi ^2\hbar u^2}{\rm Im}\left[ \eta (\omega )\right]
\int_0^\infty d\mu K_{i\mu }^2\left( \frac \omega ua\right) {\rm ch}\left[
\mu (\pi -2\gamma )\right] ,
\end{equation}
when $\varepsilon _1(\omega )=1$, $\varepsilon _2(\omega )=\varepsilon
(\omega )$, and

\begin{equation}
P(\omega )=\frac{2q^2}{\pi \hbar u^2}\left\{ {\rm Im}\frac{-1}{\varepsilon
(\omega )}\ln \frac{k_cu}\omega -\frac 2\pi {\rm Im}\left[ \frac{\eta
(\omega )}{\varepsilon (\omega )}\right] \int_0^\infty d\mu K_{i\mu
}^2\left( \frac \omega ua\right) {\rm ch}\left[ \mu (\pi -2\gamma )\right]
\right\} ,
\end{equation}
when $\varepsilon _2(\omega )=1$, $\varepsilon _1(\omega )=\varepsilon
(\omega )$. For calculation of integral in equations (21) and (22) we have
used the following expression for modified Bessel function (Bateman and
Erdelyi 1977)

\begin{equation}
K_{i\mu }^2(x)=\frac \pi {{\rm sh}(\pi \mu )}\int_0^\infty dt\sin (2\mu
t)J_0(2x{\rm sh}t),
\end{equation}
where $J_0(x)$ is the Bessel function of zeroth order. By using equation
(23) we finally find

\begin{equation}
P(\omega )=\frac{2q^2}{\pi \hbar u^2}{\rm Im}\left[ \frac{\varepsilon
(\omega )-1}{\varepsilon (\omega )+1}\right] K_0\left( 2\frac \omega
ud\right) ,
\end{equation}
when $\varepsilon _1(\omega )=1$, $\varepsilon _2(\omega )=\varepsilon
(\omega )$, and

\begin{equation}
P(\omega )=\frac{2q^2}{\pi \hbar u^2}\left\{ {\rm Im}\frac{-1}{\varepsilon
(\omega )}\ln \frac{k_cu}\omega +{\rm Im}\left[ \frac{1-\varepsilon (\omega )%
}{\varepsilon (\omega )\left( 1+\varepsilon (\omega )\right) }\right]
K_0\left( 2\frac \omega ud\right) \right\} ,
\end{equation}
when $\varepsilon _2(\omega )=1$, $\varepsilon _1(\omega )=\varepsilon
(\omega )$, $d=a\sin \gamma $ is a distance of particle from surface, $K_0$
is the modified Bessel function of zeroth order.

Dobrzynsky and Maradudin (1972) showed that in the limit of large values of $%
\mu $ or in the case of $\alpha \rightarrow \pi $ the dispersion relation of
the edge modes, Eqs (14) and (15), coincides with the dispersion relation ($%
\varepsilon =-1$) for surface plasmon modes bound to the plane interface
between a dielectric medium and vacuum. Consequently, the surface
energy-loss function in equations (16) and (19) reduces to the surface
energy-loss function in equations (24) and (25), ${\rm Im}[(\varepsilon
-1)/(\varepsilon +1)]$, in the limit $\alpha \rightarrow \pi $.

\section{Analysis and comparison with other works}

We have evaluated the excitation probabilities in Eqs (16)-(19) with the
complex dielectric function for $MgO$ taken from experimental data (Roessler
and Walker 1967), and for an 80-keV electron beam as in the experiment
(Marks 1982).

We first briefly recall the principal results of the investigations carried
out by Marks (1982) and Cowley (1982).

(i) The overall intensity in the EELS spectrum (and therefore also the
intensity of a given peak) decreases when the electron path goes from a
lateral surface to the edge of the crystal (Marks 1982) (see also figures
2-8 below).

(ii) For electron paths along a lateral surface, the intensity of a given
peak first slowly increases and then rapidly decreases exponentially as the
beam-surface distance goes from inside to outside the wedge (Marks 1982)
(see also figure 9 below).

(iii) For electron paths both parallel to the lateral surface and along the
edge, a surface plasmon at 18 eV was observed, together with a strong
enhancement of the low frequencies (in comparison with the spectrum for
electron paths through the bulk, see figure 8 below). The 18-eV peak was
attributed by Marks (1982) to a genuine surface resonance, in contrast to
Cowley's (1982) interpretation of it as due to transition radiation.

Now we are giving the numerical analyzes of the expressions (16) and (19)
for the excitation probability for $MgO$. We have utilized available bulk
optical data (Roessler and Walker 1967) to calculate the differential
energy-loss probability. Figs. 2-6 show the excitation probability for a
beam traveling in vacuum parallel to the surface of the sharp-edged wedge at
a various wedge interior and beam orientation angles $\alpha ^{*}=2\pi
-\alpha $ and $\gamma $, at a distance from edge of $a=2$ nm. In Figs. 2-4
the angle $\alpha ^{*}$ is obtuse, meanwhile in figure 5 the angle $\alpha
^{*}$ is acute and in this case the dielectric wedge is well-defined. Note
that as the angle $\alpha $ increases (the angle $\alpha ^{*}$ decreases)
and the wedge becomes well-defined, the maximum intensity of excitation
probability shifts from high value of $\omega $ to low plasmon energy
region. Also the intensity of given peak increases when the electron beam
path approaches from the symmetry plane of the wedge to the wedge lateral
surface.

Figure 6 shows the excitation probability as a function of wedge interior
angle $\alpha $ for $a=2$ nm and $\omega =13.6$ eV. The solid lines
correspond to the cases when beam position angle $\gamma $ is changed with
increasing of $\alpha $ (the following four values for angle $\gamma $ are
considered: $\gamma =\alpha /2$, $\gamma =\alpha /4$, $\gamma =\alpha /6$
and $\gamma =\alpha /10$ respectively). The dotted and dashed lines
correspond to the case when electron beam position angle is fixed with $%
\gamma =8^0$and $\gamma =\pi /2$ respectively. From this figure it follows
that excitation probability of a given mode in all curves decreases rapidly
after the value of $\alpha \simeq 3\pi /2\simeq 4.5$ rad.

The resonance at $\omega \sim 18$ eV in Figs. 3-5 and in solid curve of
figure 7 was clearly observed in the experiments (Marks 1982, Cowley 1982),
and is not so distinctly apparent in the predictions for the semi-infinite
model of the wedge (Garcia-Molina {\it et al} 1985, Marks 1982) (see also
dotted curve in figure 7). Also this resonance is absent in the case of very
large value of the wedge interior angle $\alpha ^{*}$ (see figure 2). Cowley
(1982) attributed this $\sim 18$-eV mode to transition radiation, but its
origin as a genuine surface resonance is clear from the model calculations
for the parabolically shaped wedge (Garcia-Molina {\it et al} 1985) or from
the present calculations for sharp-edged wedge.

Figure 7 shows the excitation probability for a beam travelling parallel to
the edge of the wedge and in front of it, at a distance of $a=2$ nm. The
solid curve corresponds to a well-defined wedge ($\alpha =11\pi /6$ or $%
\alpha ^{*}=\pi /6$), for an angle $\gamma =11\pi /12$, and dotted curve is
the prediction for a semi-infinite medium. This latter case, which has been
reduced by a factor of 5, is very similar to the prediction of nearly flat
parabolically shaped wedge (Garcia-Molina {\it et al} 1985).

For a wedge boundary defined by $\alpha =11\pi /6$, figure 8 shows a
comparison between the excitation probabilities of the wedge when the
electron beam passes in front of the edge ($\gamma =11\pi /12$) (dashed
curve, which is the same as solid curve in figure 7) or along one of its
lateral surfaces (solid line). In the first case $a=2$ nm, in the second one 
$a\simeq 20.1$ nm (the beam distance to the edge is taken to be 20 nm and
therefore $\gamma \simeq 0.1$). Also shown in figure 8 is the excitation
probability (reduced by a factor of 40) for a beam traveling through the
bulk of the wedge ($\alpha =\pi /6$), along its symmetry plane (dotted
line), and at a distance 20 nm from its edge. The results in figure 8 may be
compared with the experimental findings in figure 2 of Marks (1982). The
details of the experiment are reproduced by our calculations. For instance,
in the beam-lateral surface interaction spectrum, the intensity of the $\sim
18$-eV peak is greater than the intensity of the $\sim 13$-eV peak. Also,
the $\sim 18$-eV peak in this spectrum shifts to $\sim 22$ eV in the bulk
spectrum. The bulk plasmon for $MgO$ is located at $\sim 22$ eV (Roessler
and Walker 1967), as seen in figure 8. Note that for calculation of bulk
energy losses we have used the cutoff wave number $k_c=2mu/\hbar $ (Brandt 
{\it et al} 1974) which is much grater than that of used in other works ($%
k_c\sim 0.1$ nm$^{-1}$) (Illman {\it et al} 1988, Garcia-Molina {\it et al}
1985, Marks 1982).

As mentioned in (ii) above, Marks (1982) also investigated the excitation
probability of the wedge for electron beam positions ranging from $\sim 10$
nm with respect to the wedge surface, but inside the wedge, up to $\sim 10$
nm outside the wedge surface. The beam path in the experiment was far from
the edge of the wedge, and the dimensions of the cubic crystal were $\sim
100 $ nm. We have evaluated the corresponding expressions, (16) and (19),
for the wedge interior angle $\pi /3$ and for beam distances from edge $\geq
50$ nm. The results are shown in figure 9. In agreement with the
experimental results (figure 3 in the work of Marks (1982), see also point
(ii) above), the excitation probability decays exponentially with distance,
the slope being larger with increasing the energy value $\omega $. The
relative intensities of the different curves are also in agreement with the
experimental results, the curve for $\omega =10$ eV crossing the other
curves shown in figure 9. In the experiment the transition from inside to
outside the wedge is broader than figure 9 shows. One should note, however,
that our calculations was done for a point-like charged particle beam,
whereas the experimental value of the beam size was rather large, $\sim 2$
nm. We recall also that the spectrometer resolution in Marks' (1982)
experiment is 3 eV.

One of the most easily controlled variables affecting the excitation
probability is the incident-beam energy. It is clearly of practical interest
to determine the optimum incident energy which will elicit the greatest
response from a given target. To this end we have determined the energy
which would maximize the excitation probability for a given surface mode and
a wedge shape. In figure 10 we show the dependence of excitation probability
for given surface mode ($\omega =13.6$ eV) as a function of beam kinetic
energy ($1$ keV$\leq E_{{\rm kin}}\leq 100$ keV) for $\alpha =3\pi /2$ ($%
\alpha ^{*}=\pi /2$) and for various beam positions (the dotted, dashed,
solid and dot-dashed lines correspond to $\gamma =3\pi /4$, $\gamma =3\pi /8$%
, $\gamma =3\pi /16$ and $\gamma =3\pi /32$ respectively. The latter two
have been divided by factors of 2 and 6 respectively). The beam travels in
vacuum at a distance $a=2$ nm from the edge. It is evident that the
excitation probability first, increases rapidly together with the beam
energy and after some value (which grows together with $\gamma $) slowly
decreases. The plot is given for a 1-100 keV energy range and for $E_{{\rm %
kin}}<1$ keV, however, our presumption of a rectilinear trajectory becomes
questionable, and we have not shown calculations below the 1-keV level. Also
for the beam energy range $E_{{\rm kin}}>100$ keV the retardation effects
become important and a separate investigation is required.

\section{Conclusion}

We have investigated here the case of a beam traveling parallel to the edge
of a sharp-edged wedge both in vacuum and through the medium. Other
configurations, like beam trajectories at constant $z$, trajectories
intersecting the tip of the wedge, or reflecting at the lateral surface, may
also be of interest in the analysis of the experiments. Note also that the
experiment has been performed with a relatively broad probe, $\sim 2$ nm in
diameter, which is comparable to the distance from the beam to the wedge
whereas the calculations developed in this paper assume a point-like STEM
probe.

We have analyzed the electron-wedge interaction in the electrostatic limit.
The electron beam energy ($\sim 100$ keV) is large enough that one may worry
about the effect of retardation on the theoretical predictions. This is
currently being investigated.

Finally, let us mention that the expressions derived for the excitation
probability in equations (16) and (19) can be used efficiently with moderate
computing resources in practical data analysis.

\vspace{1.0in}

\begin{center}
{\bf ACKNOWLEDGMENTS}
\end{center}

It is pleasure for authors to thank Professor N.R. Arista for providing with
the optical data for a number of materials.

\vspace{1.0in}

{\bf Appendix}

Here we show that in the static limit ($u\rightarrow 0$) from equations (2)
and (4)-(9) follows the expression for the potential of a charged particle
located near the conducting ($\varepsilon \rightarrow \infty $) wedge
surface. For this we substitute $\omega =\kappa u$ in equation (2) limit of $%
u\rightarrow 0$. We find the following expression (note that inside the
conductor the potential is zero).

\begin{eqnarray}
\varphi (\rho ,\theta ,z) &=&\frac{2q}{\pi ^2}\int_0^\infty d\kappa \cos
(\kappa z)\int_{-\infty }^{+\infty }d\mu K_{i\mu }(\kappa a)K_{i\mu }(\kappa
\rho )\times  \tag{A.1} \\
&&\ \times \left\{ {\rm ch}\left[ \mu \left( \pi -\left| \theta -\gamma
\right| \right) \right] +A(\mu ){\rm ch}(\mu \theta )+B(\mu ){\rm sh}(\mu
\theta )\right\} ,  \nonumber
\end{eqnarray}
where $A(\mu )=A_\omega (\mu )$, $B(\mu )=B_\omega (\mu )$ at $\omega
\rightarrow 0$. Taking into account that for conductors $\varepsilon
\rightarrow \infty $ when $\omega \rightarrow 0$ (Landau and Lifshitz 1982)
from equations (6)-(9) we can obtain the following set of equations

\begin{equation}
A(\mu )=-{\rm ch}\left[ \mu (\pi -\gamma )\right] ,  \tag{A.2}
\end{equation}

\begin{equation}
A(\mu ){\rm ch}(\mu \alpha )+B(\mu ){\rm sh}(\mu \alpha )=-{\rm ch}\left[
\mu (\pi -\alpha +\gamma )\right] .  \tag{A.3}
\end{equation}
By solving the set of equations (A.2) and (A.3) we find

\begin{equation}
\varphi (\rho ,\theta ,z)=\frac{4q}{\pi ^2}\int_0^\infty d\kappa \cos
(\kappa z)\int_{-\infty }^{+\infty }d\mu K_{i\mu }(\kappa a)K_{i\mu }(\kappa
\rho )C(\mu ),  \tag{A.4}
\end{equation}
where

\begin{equation}
C(\mu )=\frac{{\rm sh}(\pi \mu )}{{\rm ch}(\alpha \mu )}\left\{ 
\begin{tabular}{l}
${\rm sh}(\mu \gamma ){\rm sh}[\mu (\alpha -\theta )];\qquad \gamma \leq
\theta \leq \alpha $ \\ 
${\rm sh}[\mu (\alpha -\gamma )]{\rm sh}(\mu \theta );\qquad \theta \leq
\gamma .$%
\end{tabular}
\right.  \tag{A.5}
\end{equation}
Calculating in equation (A.4) the integral by $\kappa $ (Gradshteyn and
Ryzhik 1980) we obtain the following expression for potential

\begin{eqnarray}
\varphi (\rho ,\theta ,z) &=&\frac q{\sqrt{a\rho }}\int_{-\infty }^{+\infty
}d\mu \frac{{\rm th}(\pi \mu )}{{\rm sh}(\mu \alpha )}P_{i\mu -1/2}({\rm ch}%
\eta )  \tag{A.6} \\
&&\left\{ 
\begin{tabular}{l}
${\rm sh}(\mu \gamma ){\rm sh}[\mu (\alpha -\theta )];\qquad \gamma \leq
\theta \leq \alpha $ \\ 
${\rm sh}[\mu (\alpha -\gamma )]{\rm sh}(\mu \theta );\qquad \theta \leq
\gamma ,$%
\end{tabular}
\right.  \nonumber
\end{eqnarray}
where $P_\nu (x)$ is the Legendre function of the first kind and with order $%
\nu =i\mu -1/2$,

\begin{equation}
{\rm ch}\eta =\frac{\rho ^2+a^2+z^2}{2a\rho }.  \tag{A.7}
\end{equation}
By using the known equation for the Legendre function of the second kind $%
Q_\nu (x)$ (Gradshteyn and Ryzhik 1980),

\begin{equation}
Q_{-i\mu -1/2}(z)-Q_{i\mu -1/2}(z)=\pi i{\rm th}(\pi \mu )P_{i\mu -1/2}(z), 
\tag{A.8}
\end{equation}
we can obtain

\begin{equation}
\varphi (\rho ,\theta ,z)=\frac q{\pi i\sqrt{a\rho }}\int_{-\infty
}^{+\infty }\frac{d\mu }{{\rm sh}(\mu \alpha )}Q_{-i\mu -1/2}({\rm ch}\eta
)\left\{ 
\begin{tabular}{l}
${\rm sh}(\mu \gamma ){\rm sh}[\mu (\alpha -\theta )];~\gamma \leq \theta
\leq \alpha $ \\ 
${\rm sh}[\mu (\alpha -\gamma )]{\rm sh}(\mu \theta );~\theta \leq \gamma .$%
\end{tabular}
\right.  \tag{A.9}
\end{equation}
It is convenient to use the integral presentation of Legendre function of
the second kind (Gradshteyn and Ryzhik 1980)

\begin{equation}
Q_{-i\mu -1/2}({\rm ch}\eta )=\int_\eta ^\infty \frac{d\zeta \exp (i\zeta
\mu )}{\sqrt{2({\rm ch}\zeta -{\rm ch}\eta )}}.  \tag{A.10}
\end{equation}
By using (A.10) the expression (A.9) can be written as

\begin{equation}
\varphi (\rho ,\theta ,z)=\frac q{\alpha \sqrt{2a\rho }}\int_\eta ^\infty 
\frac{d\zeta \Phi (\zeta ,\theta )}{\sqrt{{\rm ch}\zeta -{\rm ch}\eta }}, 
\tag{A.11}
\end{equation}
where

\begin{equation}
\Phi (\zeta ,\theta )=2\sum_{n=1}^\infty \exp \left( -\frac{\pi \zeta n}%
\alpha \right) \left[ \cos \left( \pi n\frac{\gamma -\theta }\alpha \right)
-\cos \left( \pi n\frac{\gamma +\theta }\alpha \right) \right] .  \tag{A.12}
\end{equation}
Finally, taking into account the expression (Gradshteyn and Ryzhik 1980)

\begin{equation}
2\sum_{n=1}^\infty \exp (-n\alpha )\cos (n\beta )=-1+\frac{{\rm sh}(\alpha )%
}{{\rm ch}(\alpha )-\cos (\beta )}  \tag{A.13}
\end{equation}
from (A.11)-(A.13) we can find the following expression for the potential
given by Landau and Lifshitz (1982)

\begin{eqnarray}
\varphi (\rho ,\theta ,z) &=&\frac q{\alpha \sqrt{2a\rho }}\int_\eta ^\infty 
\frac{d\zeta {\rm sh}\left( \frac{\pi \zeta }\alpha \right) }{\sqrt{{\rm ch}%
\zeta -{\rm ch}\eta }}\times  \tag{A.14} \\
&&\times \left\{ \frac 1{{\rm ch}\left( \frac{\pi \zeta }\alpha \right)
-\cos \left[ \frac{\pi (\theta -\gamma )}\alpha \right] }-\frac 1{{\rm ch}%
\left( \frac{\pi \zeta }\alpha \right) -\cos \left[ \frac{\pi (\theta
+\gamma )}\alpha \right] }\right\} .  \nonumber
\end{eqnarray}

\vspace{1.0in}{\bf References}

Bateman H and Erdelyi A 1977 {\it Higher Transcendental Functions }(Moscow:
Nauka)

Brandt W, Ratkowski A and Ritchie R H 1974 {\it Phys. Rev. Lett.} {\bf 33}
1325

Chen C H, Joy D C, Chen H S and Hauser J J 1986 {\it Phys. Rev. Lett.} {\bf %
57} 743

Cowley J M 1982 {\it Surf. Sci.} {\bf 114} 587

\_\_\_\_1982 {\it Phys. Rev.} B {\bf 25} 1401

Davis L D 1976 {\it Phys. Rev.} B {\bf 14} 5523

Dobrzynsky L and Maradudin A A 1972 {\it Phys. Rev.} B {\bf 6} 3810

Echenique P M, Howie A and Wheatley D J 1987 {\it Philos. Mag.} B {\bf 56}
335

Echenique P M and Pendry J B 1975 {\it J. Phys.} C {\bf 8} 2936

Ferrel T L and Echenique P M 1985 {\it Phys. Rev. Lett.} {\bf 55} 1526

Garcia-Molina R, Gras Marti A and Ritchie R H 1985 {\it Phys. Rev.} B {\bf 31%
} 121

Gradshteyn I S and Ryzhik I M 1980 {\it Table of Integrals, Series and
Products} (New York: Academic)

Illman B L, Anderson V E, Warmack R J and Ferrel T L 1988 {\it Phys. Rev.} B 
{\bf 38} 3045

Landau L D and Lifshitz E M 1982 {\it Eloctrodynamics of Continuous Media}
(Moscow: Nauka)

Marks L D 1982 {\it Solid State Commun.} {\bf 43} 727

Nersisyan H B and Hovhannisyan A V (unpublished)

Ritchie R H 1957 {\it Phys. Rev.} {\bf 106} 874

Rivacoba A, Zabala N and Echenique P M 1992 {\it Phys. Rev. Lett.} {\bf 69}
3362

Roessler D M and Walker W C 1967 {\it Phys. Rev.} {\bf 159} 733

Smythe W R 1969 {\it Static and Dynamic Electrisity} (New York: McGraw-Hill)

Walls M G and Howie A 1989 {\it Ultramicroscopy} {\bf 28} 40

Walsh C A 1989 {\it Philos. Mag.} {\bf 59} 227

Wheatley D I, Howie A and McMullan D 1983 {\it EMAG Conference Surrey}
(unpublished)

Zabala N, Rivacoba A and Echenique P M 1989 {\it Surf. Sci.} {\bf 209} 465

\begin{center}
\newpage{\bf Figure Captions}
\end{center}

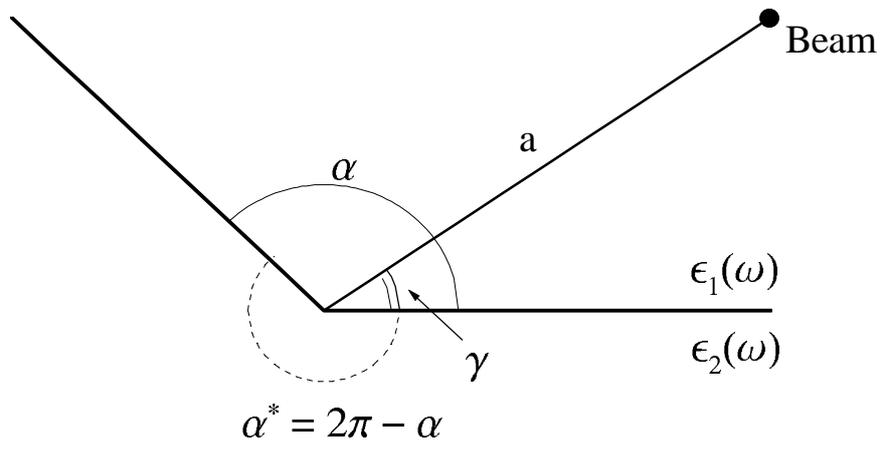
Figure 1. Dielectric wedges filling the space $-\infty <z<\infty $, $%
0<\theta <\alpha $, $\alpha <\theta <2\pi $, and characterized by an
isotropic dielectric functions $\varepsilon _1(\omega )$ and $\varepsilon
_2(\omega )$ respectively.

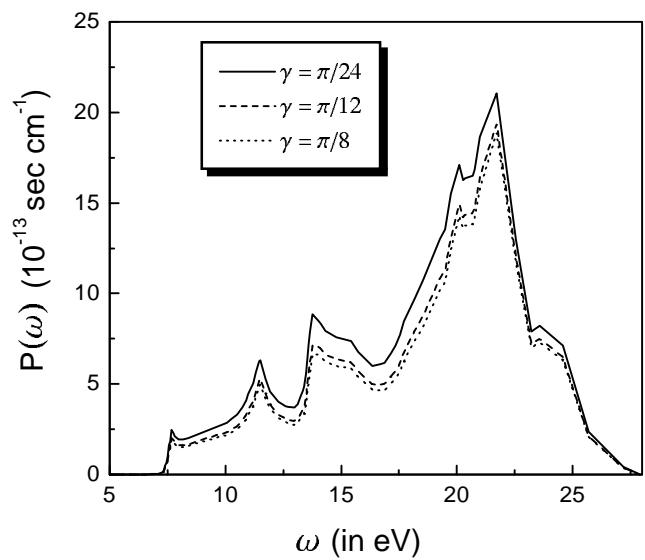
Figure 2. Excitation probability of surface modes, equation (16), for
electron beam traveling in the vacuum parallel to the surface of the wedge,
at a distance of $a=2$ nm from edge. The interior angle of the wedge is
obtuse ($\alpha =\pi /4$ or $\alpha ^{*}=7\pi /4$). The solid, dashed and
dotted lines correspond to the three positions of the beam $\gamma =\pi /24$%
, $\gamma =\pi /12$ and $\gamma =\pi /8$ respectively. The electron beam
energy is 80 keV.

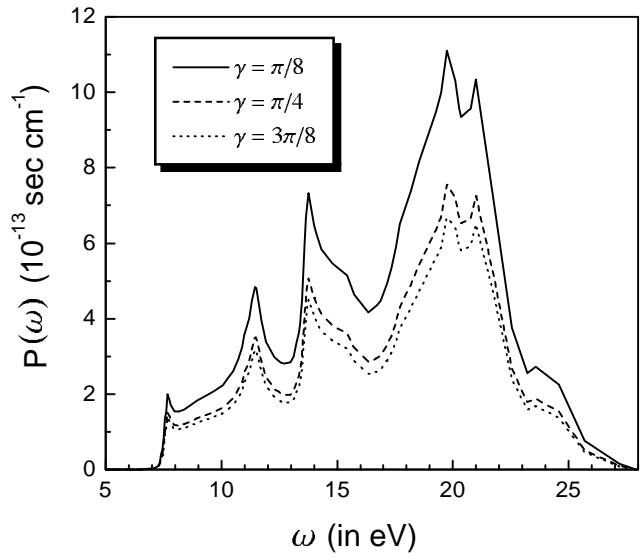
Figure 3. Same as in figure 2, but here $\alpha =3\pi /4$ (or $\alpha
^{*}=5\pi /4$). The solid, dashed and dotted lines correspond to the three
positions of the beam $\gamma =\pi /8$, $\gamma =\pi /4$ and $\gamma =3\pi
/8 $ respectively.

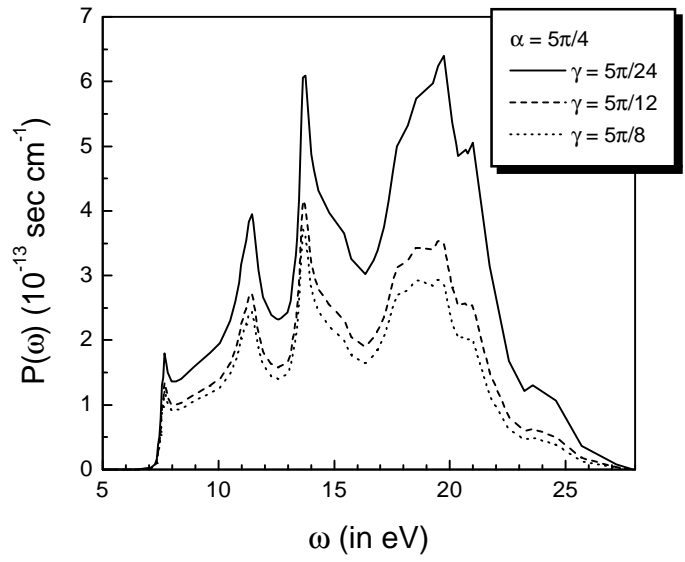
Figure 4. Same as in figure 2, but here $\alpha =5\pi /4$ (or $\alpha
^{*}=3\pi /4$). The solid, dashed and dotted lines correspond to the three
positions of the beam $\gamma =5\pi /24$, $\gamma =5\pi /12$ and $\gamma
=5\pi /8$ respectively.

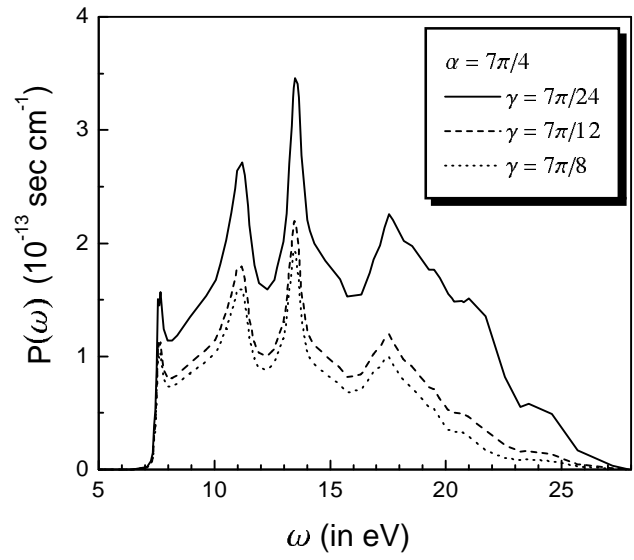
Figure 5. Same as in figure 2, but here $\alpha =7\pi /4$ (or $\alpha
^{*}=\pi /4$, i.e. the wedge is acute-angled and well-defined). The solid,
dashed and dotted lines correspond to the three positions of the beam $%
\gamma =7\pi /24$, $\gamma =7\pi /12$ and $\gamma =7\pi /8$ respectively.

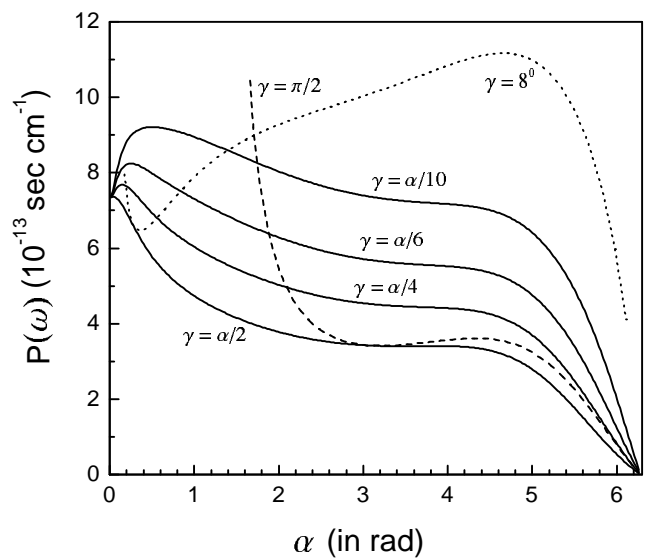
Figure 6. Excitation probability of given surface mode $\omega =13.6$ eV as
a function of $\alpha $ (in radian) for electron beam traveling in the
vacuum parallel to the surface of the wedge at a distance of $a=2$ nm from
edge. The solid lines correspond to the varying beam position angles ($%
\gamma =\alpha /2$, $\gamma =\alpha /4$, $\gamma =\alpha /6$ and $\gamma
=\alpha /10$ respectively). The dotted and dashed lines correspond to the
fixed values of beam position angles $\gamma =8^0$ and $\gamma =\pi /2$
respectively. The electron beam energy is 80 keV.

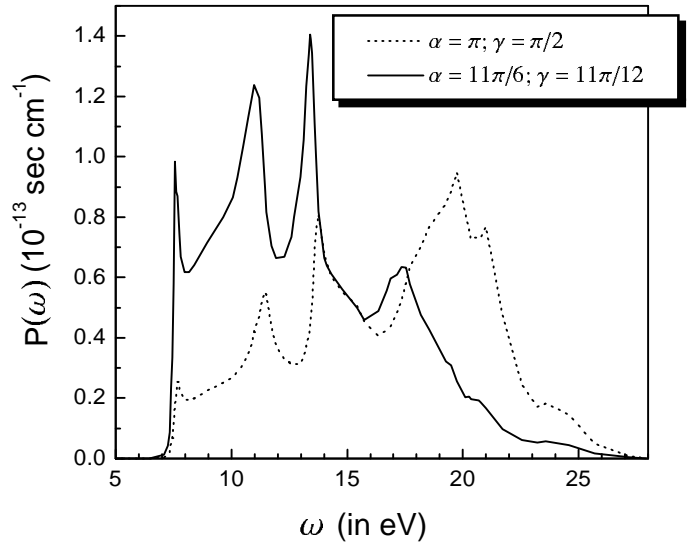
Figure 7. Excitation probability of surface modes, equation (16), for
electron beam traveling in the vacuum parallel to the surface of the wedge
and in front of it, at a distance of $a=2$ nm from edge. The solid line
corresponds to the well-defined wedge ($\alpha =11\pi /6$ or $\alpha
^{*}=\pi /6$) with the beam position angle $\gamma =11\pi /12$, the dotted
line corresponds to the semi-infinite wedge ($\alpha =\pi $, $\gamma =\pi /2$%
). In the latest case the probability has been divided by a factor of 5. The
electron beam energy is 80 keV.

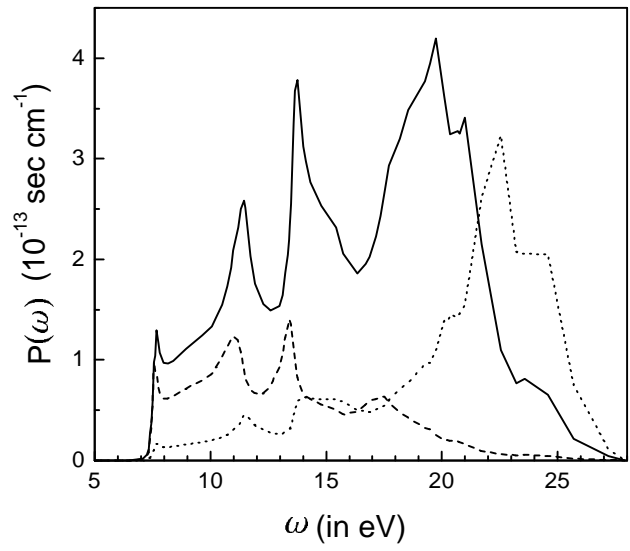
Figure 8. Excitation probabilities of surface and bulk modes, equations (16)
and (19), for electron beam traveling in the vacuum along the edge of the
well-defined wedge ($\alpha =11\pi /6$ or $\alpha ^{*}=\pi /6$), at a
distance of 2 nm from the edge and with angular position $\gamma =11\pi /12$
(dashed line). This curve is same as the solid curve in figure 7; along a
lateral surface, at a distance of 2 nm from it and in this case $\alpha
=11\pi /6$ ($\alpha ^{*}=\pi /6$), $a\simeq 20.1$ nm ($\gamma \simeq 0.1$)
(solid line); and dotted line through the bulk of the wedge (equation (19)),
along the symmetry plane ($\alpha =\pi /6$, $\gamma =\pi /12$) and at a
distance of $a=20$ nm from the edge. The spectrum in dotted curve has been
divided by a factor of 40. The electron beam energy is 80 keV.

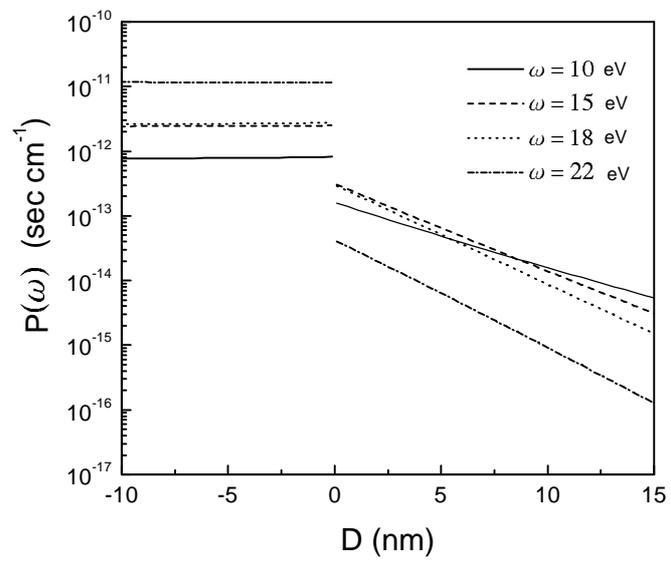
Figure 9. Excitation probability, for given modes $\omega $ (in eV),
equations (16) and (19), for electron beam paths at a varying distance $D$
(in nm) from the lateral surface of the wedge. The beam paths are far away
from the edge ($\geq 50$ nm), and range from inside ($D<0$) to outside ($D>0$%
) the wedge. Wedge interior angle in both cases is $\pi /3$. The electron
beam energy is 80 keV.

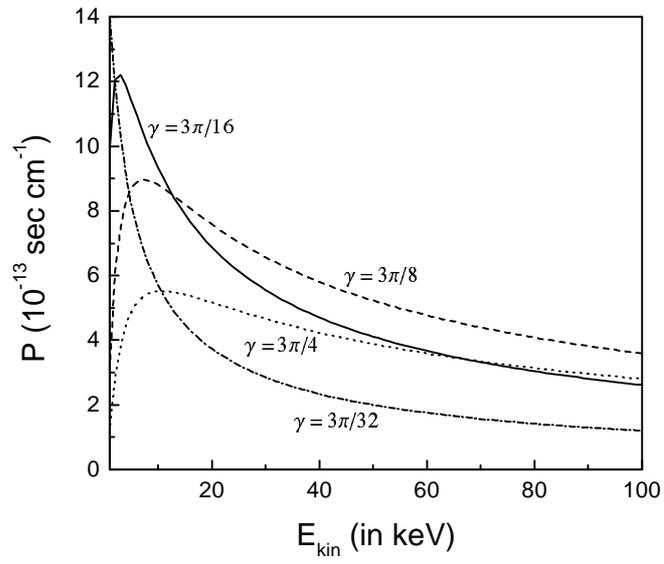
Figure 10. Excitation probability for given surface mode $\omega =13.6$ eV
as a function of electron beam kinetic energy $E_{{\rm kin}}$ ($1$ keV$\leq
E_{{\rm kin}}\leq 100$ keV). The beam moves in the vacuum parallel to the
surface of the well-defined wedge ($\alpha =3\pi /2$ or $\alpha ^{*}=\pi /2$%
), at a distance of 2 nm from the edge. The dotted, dashed, solid and
dot-dashed lines correspond to $\gamma =3\pi /4$, $\gamma =3\pi /8$, $\gamma
=3\pi /16$ and $\gamma =3\pi /32$ respectively. The probabilities in the
curves with $\gamma =3\pi /16$ and $\gamma =3\pi /32$ have been divided by
factors of 2 and 6 respectively.

\end{document}